\documentclass[%
 superscriptaddress,
 preprint,
 showkeys
 amsmath,amssymb,
 prx,
]{revtex4-2}

\usepackage[version=4]{mhchem}
\usepackage{graphicx}% Include figure files
\usepackage{dcolumn}% Align table columns on decimal point
\usepackage{bm}% bold math
\usepackage{setspace}
\usepackage{caption}
\usepackage{upgreek} 
\begin{document}

\title{Materializing Rival Ground States in the Barlowite Family of Kagome Magnets: Quantum Spin Liquid, Spin Ordered, and Valence Bond Crystal States}

\author{Rebecca W. Smaha}
 \thanks{These authors contributed equally to the work}
 \email{rsmaha@stanford.edu}
 \affiliation{Stanford Institute for Materials and Energy Sciences, SLAC National\\Accelerator Laboratory, Menlo Park, California 94025, USA}
 \affiliation{Department of Chemistry, Stanford University, Stanford, California 94305, USA}
\author{Wei He}
 \thanks{These authors contributed equally to the work.}
 \affiliation{Stanford Institute for Materials and Energy Sciences, SLAC National\\Accelerator Laboratory, Menlo Park, California 94025, USA}
 \affiliation{Department of Materials Science and Engineering, Stanford University, Stanford, California 94305, USA} 
\author{Jack Mingde Jiang}
 \affiliation{Stanford Institute for Materials and Energy Sciences, SLAC National\\Accelerator Laboratory, Menlo Park, California 94025, USA}
 \affiliation{Department of Applied Physics, Stanford University, Stanford, California 94305, USA}
\author{Jiajia Wen}
 \affiliation{Stanford Institute for Materials and Energy Sciences, SLAC National\\Accelerator Laboratory, Menlo Park, California 94025, USA}
\author{Yi-Fan Jiang}
 \affiliation{Stanford Institute for Materials and Energy Sciences, SLAC National\\Accelerator Laboratory, Menlo Park, California 94025, USA}
\author{John P. Sheckelton}
 \affiliation{Stanford Institute for Materials and Energy Sciences, SLAC National\\Accelerator Laboratory, Menlo Park, California 94025, USA}
\author{Charles J. Titus}
 \affiliation{Department of Physics, Stanford University, Stanford, California 94305, USA}
 \author{Suyin Grass Wang}
 \affiliation{NSF's ChemMatCARS, Center for Advanced Radiation Sources, c/o Advanced\\Photon Source/ANL, The University of Chicago, Argonne, Illinois 60439, USA}
\author{Yu-Sheng Chen}
 \affiliation{NSF's ChemMatCARS, Center for Advanced Radiation Sources, c/o Advanced\\Photon Source/ANL, The University of Chicago, Argonne, Illinois 60439, USA}
\author{Simon J. Teat}
 \affiliation{Advanced Light Source, Lawrence Berkeley National Laboratory, Berkeley, California 94720, USA}
\author{Adam A. Aczel}
 \affiliation{Neutron Scattering Division, Oak Ridge National Laboratory, Oak Ridge, Tennessee 37831, USA}
 \affiliation{Department of Physics and Astronomy, University of Tennessee, Knoxville, Tennessee 37996, USA}
\author{Yang Zhao}
 \affiliation{NIST Center for Neutron Research, National Institute of Standards and\\Technology, Gaithersburg, MD 20899-6102 USA}
 \affiliation{Department of Materials Science and Engineering, University of Maryland, College Park, Maryland 20742, USA}
\author{Guangyong Xu}
 \affiliation{NIST Center for Neutron Research, National Institute of Standards and\\Technology, Gaithersburg, MD 20899-6102 USA}
\author{Jeffrey W. Lynn}
 \affiliation{NIST Center for Neutron Research, National Institute of Standards and\\Technology, Gaithersburg, MD 20899-6102 USA}
\author{Hong-Chen Jiang}
 \affiliation{Stanford Institute for Materials and Energy Sciences, SLAC National\\Accelerator Laboratory, Menlo Park, California 94025, USA}
\author{Young S. Lee}
 \email{youngsl@stanford.edu}
 \affiliation{Stanford Institute for Materials and Energy Sciences, SLAC National\\Accelerator Laboratory, Menlo Park, California 94025, USA}
 \affiliation{Department of Applied Physics, Stanford University, Stanford, California 94305, USA}

\begin{abstract}
The spin-$\frac{1}{2}$ kagome antiferromagnet is considered an ideal host for a quantum spin liquid ground state. We find that when the bonds of the kagome lattice are modulated with a periodic pattern, new quantum ground states emerge. Newly synthesized crystalline barlowite (\ce{Cu4(OH)6FBr}) and Zn-substituted barlowite demonstrate the delicate interplay between singlet states and spin order on the spin-$\frac{1}{2}$ kagome lattice. Comprehensive structural measurements demonstrate that our new variant of barlowite maintains hexagonal symmetry at low temperatures with an arrangement of distorted and undistorted kagome triangles, for which numerical simulations predict a pinwheel valence bond crystal (VBC) state instead of a quantum spin liquid (QSL). The presence of interlayer spins eventually leads to an interesting pinwheel \textit{q=0} magnetic order. Partially Zn-substituted barlowite (\ce{Cu$_{3.44}$Zn$_{0.56}$(OH)6FBr}) has an ideal kagome lattice and shows QSL behavior, indicating a surprising robustness of the QSL against interlayer impurities. The magnetic susceptibility is similar to that of herbertsmithite, even though the \ce{Cu$^{2+}$} impurities are above the percolation threshold for the interlayer lattice and they couple more strongly to the nearest kagome moment. This system is a unique playground displaying QSL, VBC, and spin order, furthering our understanding of these highly competitive quantum states. 
\end{abstract}

%Use showkeys class option if keyword display desired
\maketitle

\begin{figure*}[htbp]
\includegraphics[width=14cm]{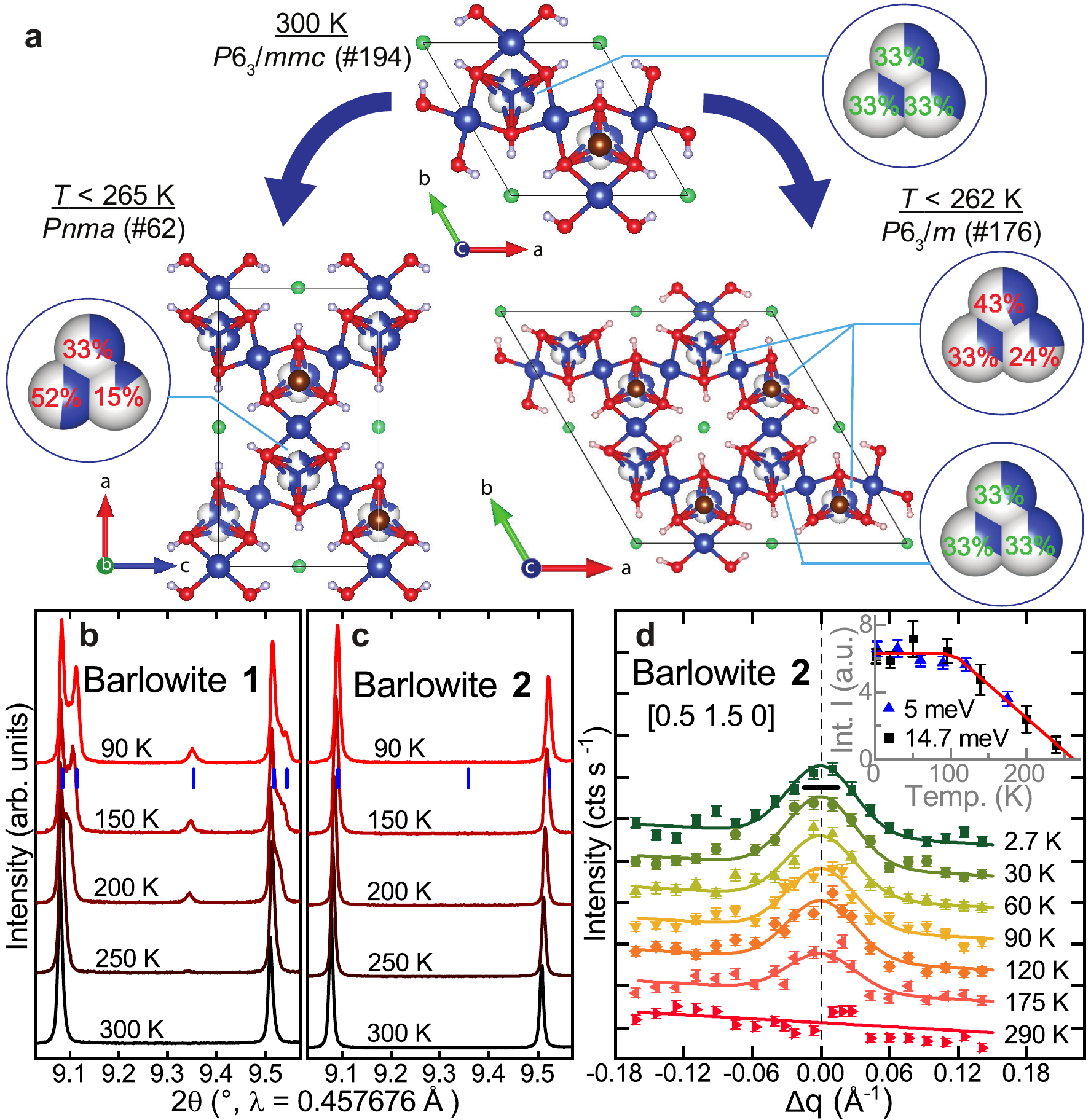}
\caption{\textbf{Structures of barlowite \textbf{1} and \textbf{2}.} \textbf{a} Schematic showing the two phase transitions of barlowite \textbf{1} and \textbf{2}. The interlayer occupancy values are taken from SCXRD (Supplementary Table 3). \textbf{b} PXRD patterns from $T = 90$ K to $300$ K showing a \textit{Pnma} superlattice peak and orthorhombic splitting in barlowite \textbf{1}. \textbf{c} PXRD patterns from $T = 90$ K to $300$ K showing no visible \textit{P}6$_3$/\textit{m} superlattice peak nor orthorhombic splitting in barlowite \textbf{2}. Bragg reflections of \textit{Pnma} and \textit{P}6$_3$/\textit{m}, respectively, at $T = 90$ K are indicated by blue tick marks. \textbf{d} Elastic neutron scattering on single crystalline barlowite \textbf{2} with $E_\mathrm{i}=5$ meV showing the temperature dependence of the \textit{P}6$_3$/\textit{m} superlattice peak [0.5 1.5 0]. The notation of the high-temperature space group \textit{P}6$_3$/\textit{mmc} is used to index peaks. Inset: temperature dependence of the integrated intensity measured along the longitudinal and transverse directions of the wavevector $\vec{Q}$ in the scattering plane with $E_\mathrm{i}=5$ and $E_\mathrm{i}=14.7$ meV, respectively, indicating that the structural transition occurs at $T =262(8)$ K. Uncertainties are statistical in origin and represent one standard deviation.}
\label{fgr:1}
\end{figure*}

%\tableofcontents
\begin{flushleft}
\section{\label{sec:Intro}Introduction}

Identifying the ground state for interacting quantum spins on the kagome lattice is an important unresolved question in condensed matter physics owing to the great difficulty in selecting amongst competing states that are very close in energy. Antiferromagnetic (AF) spins on this lattice are highly frustrated, and for spin $S=\frac{1}{2}$ systems the ground state does not achieve magnetic order and is believed to be a quantum spin liquid (QSL).\cite{Sachdev1992,Ran2007,Hermele2008,Jiang2008,Yan2011,Depenbrock2012,Jiang2012,Hao2013,He2017} The QSL is an unusual magnetic ground state, characterized by long-range quantum entanglement of the spins with the absence of long-range magnetic order.\cite{Balents2010,Norman2016,Mendels2016,Savary2017} The recent identification of herbertsmithite (\ce{Cu3Zn(OH)6Cl2})\cite{Braithwaite2004,Shores2005,Han2012,Fu2015} as a leading candidate QSL material has ignited intense interest in further understanding similar kagome materials.

For the ideal $S=\frac{1}{2}$ kagome nearest neighbor Heisenberg model, theoretical calculations point towards various QSL ground states (gapped and ungapped), though the exact ground state is still a matter of great debate.\cite{Sachdev1992,Ran2007,Hermele2008,Jiang2008,Yan2011,Depenbrock2012,Jiang2012,Hao2013,Zhu2015,He2017,Mei2017,Liao2017} Moreover, early calculations indicated that a valence bond crystal (VBC) state is a close rival to the QSL.\cite{Hastings2000,Syromyatnikov2002,Nikolic2003,Singh2007} However, real kagome materials often have additional interactions that relieve the frustration and drive the moments to magnetically order, \cite{Kim2008,Yoshida2012} indicating that the ground state depends sensitively on small perturbations and lattice distortions. Here, we present a material system in which small changes can be made to the kagome lattice, thereby revealing closely related QSL, VBC, and spin ordered states. Importantly, we make close connection with theory by performing numerical simulations for the quantum moments based on the specific symmetry of the material.

In herbertsmithite, the kagome layers are separated by non-magnetic \ce{Zn$^{2+}$} ions, which preserves the two-dimensional character of the magnetic layers. The kagome layers stack in an ABC sequence. While \ce{Zn$^{2+}$} does not mix onto the highly Jahn-Teller distorted \ce{Cu$^{2+}$} kagome sites, up to 15\% \ce{Cu$^{2+}$} can mix onto the interlayer \ce{Zn$^{2+}$} sites,\cite{Freedman2010} which complicates the observations of intrinsic QSL behavior.\cite{Han2016b} Barlowite (\ce{Cu4(OH)6FBr}) is another recently discovered mineral with layered \ce{Cu$^{2+}$} kagome planes that are stacked in an AA sequence, different from herbertsmithite.\cite{Elliott2014} The \ce{Cu$^{2+}$} moments between the kagome layers cause long-range magnetic order at $T_\mathrm{N} \approx10$--$15$ K, much lower than the Curie-Weiss temperature of  $\Theta \approx-130$ K.\cite{Han2014,Jeschke2015,Han2016,Smaha2018} The magnetic structure of orthorhombic barlowite has been interpreted using conflicting models based on neutron powder diffraction data.\cite{Feng2018,Tustain2018} The precise determination of the magnetic order is made challenging by the weak cross-section of $S=\frac{1}{2}$ \ce{Cu$^{2+}$} moments, and the stronger Bragg peaks available from single crystal measurements would be an important improvement.

While barlowite itself is an intriguing quantum magnet, it is also the parent of a QSL candidate, Zn-substituted barlowite, whose stability was first predicted by first principles calculations.\cite{Liu2015a,Guterding2016a} Substituting barlowite's interlayer site with \ce{Zn$^{2+}$} isolates the \ce{Cu$^{2+}$} kagome layers, leading to \ce{Cu3Zn$_{x}$Cu$_{1-x}$(OH)6FBr}. Calculations also predict significantly fewer magnetic \ce{Cu$^{2+}$} ``impurities'' on the interlayer site than in herbertsmithite, making it an attractive material to study.\cite{Liu2015a}  Polycrystalline samples show no magnetic order, and we find its Curie-Weiss temperature to be as high as $\Theta = -253$ K.  A gapped ground state with Z$_2$ topological order has been proposed.\cite{Feng2017,Wei2017,Feng2018} The first single crystals of Zn-substituted barlowite were recently synthesized with $x = 0.33$, which suppresses magnetic order to $T = 4$ K.\cite{Smaha2018} Our new synthesis techniques produce large single crystals of barlowite and Zn-substituted barlowite, and measurements of the magnetic and structural properties allow for detailed studies of how the QSL tendencies on the $S=\frac{1}{2}$ kagome lattice are affected by subtle structural symmetry-lowering and magnetic impurities on the interlayer sites. 

\section{Results}

\subsection{\label{sec:barlowiteStructure}New high-symmetry barlowite: a pinwheel modulated kagome lattice}

The low-temperature symmetry of the previously-studied barlowite (denoted barlowite \textbf{1}) has been subject to differing interpretations.\cite{Han2014,Jeschke2015,Han2016,Pasco2018,Feng2018,Tustain2018,Henderson2018} We performed a comprehensive structural investigation employing a combination of high resolution synchrotron powder and single crystal X-ray diffraction (PXRD and SCXRD) as well as neutron powder diffraction (NPD) to definitively determine the structure of barlowite \textbf{1}, which transforms to orthorhombic \textit{Pnma} below $T \approx 265$ K (Supplementary Figures 2--4; Supplementary Tables 2--3, 5--8, and 10--11). Figure \ref{fgr:1}b shows the emergence of superlattice peaks and orthorhombic peak splitting related to this transition. We have also grown sizable single crystals of a different variant of barlowite using a new chemical reaction.\cite{Smaha2018} Intriguingly, these crystalline samples (denoted barlowite \textbf{2}) have a low-temperature structure distinct from that of barlowite \textbf{1}, belonging to the higher-symmetry hexagonal space group \textit{P}6$_3$/\textit{m}. X-ray precession images (Supplementary Figure 1) exhibit the emergence of superlattice peaks consistent with this space group, and no orthorhombic splitting is observed (Figure \ref{fgr:1}c, Supplementary Figures 5--6, and Supplementary Table 9)--- ruling out any orthorhombic distortion down to 20 times smaller than in barlowite \textbf{1}. In the following, all peaks are indexed according to the high-temperature space group (\textit{P}6$_3$/\textit{mmc}). Figure \ref{fgr:1}d shows a superlattice peak measured via elastic neutron scattering in a single crystal sample of barlowite \textbf{2}. The inset shows the temperature dependence of the integrated intensity from both longitudinal and transverse scans (Figure \ref{fgr:1}d and Supplementary Figure 20, respectively), indicating that the structural transition occurs at $T \approx 262(8)$ K. The fitted width of the superlattice peak (Supplementary Table 17) indicates an in-plane domain size of $\sim61$ {\AA} assuming a finite-size domain model.\cite{x-ray}

In both variants of barlowite, the structural phase transition is characterized by changes in the relative occupancies of the interlayer \ce{Cu$^{2+}$} site, summarized in Figure \ref{fgr:1}a. At $T = 300$ K, the interlayer \ce{Cu$^{2+}$} is disordered with equal occupancy over three symmetry-equivalent sites with distorted trigonal prismatic coordination (point group $C_{2v}$).  Below the phase transition in barlowite \textbf{1}, we find that the interlayer \ce{Cu$^{2+}$}s become disordered over three symmetry-inequivalent sites with unequal occupancies (Figure \ref{fgr:1}a). The much higher resolution of our new synchrotron X-ray diffraction data compared to neutron powder diffraction may explain why these occupancies have not been reported in previous structural models.\cite{Feng2018,Pasco2018,Tustain2018} For our fit results, see Supplementary Tables 3 and 8. In barlowite \textbf{2}, the superlattice unit cell is doubled along the \textit{a} and \textit{b} axes. As shown in Figure \ref{fgr:1}A, two of the eight interlayer \ce{Cu$^{2+}$} locations have equal occupancy of the three sites, while the remaining six \ce{Cu$^{2+}$} locations have unequal occupancies. Barlowite \textbf{2} has a much less drastic difference in relative occupancy compared to barlowite \textbf{1}, reflecting higher similarity with the high-temperature structure and less symmetry breaking.

While kagome materials often undergo structural transitions upon cooling that relieve the frustration,\cite{Reisinger2011,Kato2001} in barlowite the kagome layer remains substantially intact throughout the transition. At room temperature, there is only one kagome \ce{Cu$^{2+}$} site, forming equilateral triangles on the ideal kagome lattice. In both lower symmetry space groups, there are two distinct kagome \ce{Cu$^{2+}$} sites; thus, there are slight variations in the Cu--Cu distances---on the order of 0.01 {\AA} (see Supplementary Table 1). In barlowite \textbf{1}, all kagome triangles are distorted, with the Cu--O--Cu kagome bond angles varying by $\pm1.2^{\circ}$.  In barlowite \textbf{2}, two kagome equilateral triangles exist in the supercell with identical Cu--O--Cu kagome bond angles, and the other six triangles are only slightly distorted with bond angles varying by $\pm0.8^{\circ}$ around their triangles. In principle, these slight changes in bond lengths and bond angles should lead to different magnitudes of the magnetic exchange interaction, though the differences may be small. Most interestingly, the bonds form a pinwheel-type pattern, as depicted in Figure \ref{fgr:table}. Also depicted are the bond patterns in barlowite \textbf{1}, which has a strong orthorhombic distortion, and also in Zn-substituted barlowite (\ce{Cu$_{4-x}$Zn$_{x}$(OH)6FBr} with $x>0.5$), which has the ideal kagome lattice.

\subsection{\label{sec:DMRG}Understanding the ground state physics with numerical simulations}

To explore the possible magnetic ground states stemming from these distinct structures, we investigated the ground state properties of the distorted Heisenberg model with Hamiltonian $H=\sum_{\langle ij \rangle} J_{ij} {\bf S}_i \cdot {\bf S}_j$ on the kagome lattice using the density-matrix renormalization group (DMRG) technique.\cite{White1992} Our results for the isotropic case ($J_{ij}=J$) are consistent with the ground state of the system being a QSL (Figure \ref{fgr:3}c), as expected. We further calculated the model with couplings extracted from barlowite \textbf{1} and barlowite \textbf{2}, with spatial anisotropy $J'<J$ as illustrated in Figure \ref{fgr:3}. While the longer range and interlayer magnetic couplings are neglected, as well as the Dzyaloshinskii-Moriya (DM) interaction and exchange anisotropy, due to numerical limitations, the simplified model captures essential features of the magnetic correlations in the kagome layer.

To make predictions for experimental measurements of the magnetic correlations, the equal-time spin structure factor $S(q)=\frac{1}{N}\sum_{i,j} \left\langle {\bf S}_i \cdot{\bf S}_j \right\rangle e^{-iq(r_i-r_j)}$ is calculated. As shown in Figure \ref{fgr:3}a, the simplified model for barlowite \textbf{1} yields a very sharp peak at [1 1 0], implying a strong magnetic ordering tendency. In contrast, the simplified model for barlowite \textbf{2} only exhibits a very broad peak (Figure \ref{fgr:3}b), indicating gapped spin excitations with exponentially decaying spin-spin correlations. The pinwheel singlet pattern is consistent with a 12-site valence bond crystal (VBC) state, which is a close rival to the QSL state proposed in early calculations.\cite{Hastings2000,Syromyatnikov2002,Nikolic2003,Singh2007} In Figure \ref{fgr:3}d, we show the phase diagram of the $J$-$J'$ Heisenberg model on the kagome lattice as a function of $J'/J$, where a phase transition between the pinwheel VBC order and the QSL state is found around $J'\sim 0.998(1)J$. The nature of the phase transition observed here is first order (see Supplementary Figure 25); this is similar to the transition between the QSL and a 36-site VBC demonstrated in a previous study.\cite{Yan2011} That this transition occurs so close to $J'/J=1$ clearly demonstrates the close proximity of the ground state energies of the pinwheel VBC and the QSL. In addition, the calculated value for the spin gap of the pinwheel VBC is $\sim 0.15J$; this was calculated deeper in the VBC phase at $J'/J=0.95$.

The model's prediction that barlowite \textbf{1} has a spin-ordered ground state is borne out by experimental results.\cite{Han2014,Jeschke2015,Han2016,Smaha2018} While we do not observe long-range magnetic order in the kagome plane, it may be induced by ferromagnetic (FM) coupling to interlayer spins.\cite{Jeschke2013} On the other hand, the model for barlowite \textbf{2} predicts a VBC ground state, although the model neglects the coupling to interlayer spins---which undoubtedly play a role. We also neglect the effects of the DM interaction and exchange anisotropy, which exist in similar materials,\cite{Han2012b} but the values for these interactions in barlowite have not yet been determined. Therefore, barlowite \textbf{1} and \textbf{2} should have substantially different phase transitions since the preferred ground states on the kagome planes have different symmetries, as we explore below.

\subsection{\label{sec:barlowiteMagnetism}Quantum magnetism of barlowite \textbf{2}: accommodating the pinwheel modulation}

Magnetic susceptibility measurements on barlowite \textbf{1} and \textbf{2} (Figure \ref{fgr:barl}a) illustrate the significant differences between the two variants. While barlowite \textbf{1} has a transition at $T_\mathrm{N} \approx 16$ K, barlowite \textbf{2} has two transitions: a broad transition at $T_\mathrm{N1} \approx10$ K and a second, sharp transition at $T_\mathrm{N2} \approx6$ K. The heat capacity and entropy related to the magnetic transitions ($C_\mathrm{mag}$ and $S$) of single crystalline barlowite \textbf{2} have a gradual onset of short-range order that clearly becomes significant well below $T = 16$ K (Figure \ref{fgr:barl}d and Supplementary Figure 16). For both \textbf{1} and \textbf{2}, the entropy change associated with the magnetic transition is approximately 1/4 of the expected value for the four \ce{Cu$^{2+}$}s in the formula unit, implying that the interlayer \ce{Cu$^{2+}$} precipitates the order.\cite{Smaha2018} In an applied field, the transition broadens and shifts to lower temperature with a crossover at $T \approx 10$ K, similar to the $T_\mathrm{N1}$ of \textbf{2}. 

Our single crystals of barlowite \textbf{2} are large enough for AC susceptibility ($\chi$') and DC magnetization ($M$) measurements, revealing significant anisotropy (Figures \ref{fgr:barl}b and c; Supplementary Figures 11--12). A net FM moment arises in the \textit{ab} plane, while the susceptibility along the \textit{c}-axis remains suppressed through the two magnetic transitions. The magnetization shows a similar story: at $T = 2$ K, there is hysteresis when $\mu_\mathrm{0}H$ is parallel to the \textit{ab} plane but not when $\mu_\mathrm{0}H$ is along the \textit{c}-axis. At $T = 20$ K, above the magnetic transitions, the two orientations are nearly identical. While the overall interactions in the kagome layers are AF (Curie-Weiss temperature $\Theta = -114(1)$ K; see Supplementary Table 15), there is a net FM moment of $\approx0.11 \upmu_\mathrm{B}$ per formula unit in the magnetically ordered, low-temperature state, obtained as the height of the hysteresis loop. The FM moment is $\sim 3$ times smaller than the net moment of barlowite \textbf{1} (Figure \ref{fgr:barl}c), indicating a more fragile spin ordered state.

To directly determine the magnetic structure, we co-aligned 1,115 deuterated crystals for magnetic neutron diffraction measurements. A large sample is required since the ordered moment is small and the measured magnetic Bragg peaks are weak. Strikingly, half-integer and integer magnetic Bragg peaks exhibit distinct behavior. Figure \ref{fgr:4}a shows a comparison of longitudinal scans through the [0.5 0.5 0] and [1 0 0] magnetic Bragg peaks. Half-integer peaks are noticeably broader than integer peaks and have enhanced background scattering (see Supplementary Figure 21 and Supplementary Table 17). After deconvolving the instrumental resolution, the typical domain size extracted using a finite-size domain model\cite{x-ray} for the half-integer peaks is $\sim118$ {\AA}, while the integer peaks yield a much larger domain size $>360$ {\AA} (close to the limit of instrument resolution). Their peak intensities also have distinct temperature dependences below $T \approx 10$ K, the $T_\mathrm{N1}$ observed in AC susceptibility  (Figure \ref{fgr:4}b and Supplementary Figure 19). The [0.5 0.5 0] peak increases smoothly below $T \approx 10$ K, matching $M^2$ measured via magnetization. However, the integer peaks are extremely weak in the intermediate temperature range $6$ K $<T<10$ K and only rise rapidly below $T \approx 6$ K ($T_\mathrm{N2}$) as resolution-limited peaks. This indicates that the two transitions denote the ordering of two sets of moments, which we identify with interlayer moments that have short-range order below $T_\mathrm{N1} \approx 10$ K and kagome moments that have long-range order below $T_\mathrm{N2} \approx 6$ K. This is consistent with the magnetic structure factor calculations that we discuss below.

A simple spin configuration based on the structural symmetry is a variation of the \textit{q=0} arrangement for the kagome \ce{Cu$^{2+}$}s in which the spins on each triangle are oriented 120$^{\circ}$ to each other (Figure \ref{fgr:4}c; Supplementary Figures 22--23; Supplementary Tables 18--22). Since there are two crystallographically distinct kagome \ce{Cu$^{2+}$} sites---the sets of red and green spins---we allowed them to have different moment sizes and different in-plane rotation directions. The best magnetic structure based on these assumptions is depicted in Figure \ref{fgr:4}c. Each set of distinct kagome spins has a local \textit{q=0} arrangement with a relative angle of $\sim28^{\circ}$ between them (a relative angle of zero would denote a perfect \textit{q=0} arrangement). We call this a ``pinwheel \textit{q=0}'' state. In addition, parallel lines of spin in the kagome plane have an out-of-plane rotation angle of $\sim29^{\circ}$ (determined by the intensity of the [1 1 1] magnetic Bragg peak). This out-of-plane tilting along lines is consistent with the allowed zero energy modes in a \textit{q=0} structure with an easy-axis exchange anisotropy (here, the $C_3$ rotational symmetry of the lattice is spontaneously broken). The net FM moment is 0.033 $\upmu_\mathrm{B}$ per formula unit, which is smaller than the FM moment of 0.11 $\upmu_\mathrm{B}$ per formula unit obtained from the magnetization; the difference may be related to the field-cooled protocol used to determine the FM moment. The magnetic peaks of barlowite \textbf{2} are considerably weaker than those observed in barlowite \textbf{1},\cite{Feng2018,Tustain2018} supporting its closer proximity to the QSL ground state.

While seemingly complicated, this magnetic structure is consistent with the underlying microscopic magnetic interactions. \textit{Ab initio} calculations indicate a strong FM interaction between the interlayer spin and its NN kagome spin.\cite{Jeschke2015} This FM interaction would compete with the AF interactions between kagome spins and perturb the perfect \textit{q=0} configuration in the manner that we observe. We calculated the energy of various spin configurations including the AF $J$ between kagome moments and the $J_{FM}$ between the interlayer and kagome moments. Based on our spin structure, the estimated FM couplings between the interlayer and kagome spins are indeed strong ($-0.89J$ on average; see Supplementary Figure 24), consistent with the previous calculations.\cite{Jeschke2015}  We find that the same microscopic rules are consistent with the spin structure for barlowite \textbf{1}\cite{Tustain2018} (Figure \ref{fgr:4}d), albeit with an orthorhombic magnetic configuration. For completeness, we also fit barlowite \textbf{2} with the orthorhombic spin structure proposed for barlowite \textbf{1}\cite{Tustain2018} (Supplementary Tables 23--25); however, this model is not obviously compatible with the underlying crystal structure and interactions of barlowite \textbf{2}. We note that the weak scattering from $S=\frac{1}{2}$ moments limits the number of magnetic Bragg peaks observable in the single crystal sample, so that the proposed spin structure, while consistent with the structural symmetry and interactions, is not necessarily unique.

We interpret the transitions at $T_\mathrm{N1}=10$ K and $T_\mathrm{N2}=6$ K in terms of a successive ordering of different spins. First, the interlayer spins begin ordering below $T_\mathrm{N1}$ since these moments mainly contribute to the half-integer magnetic peaks. The peak width indicates that they order with a short correlation length, consistent with the lack of a sharp anomaly in the heat capacity. This is followed at $T_\mathrm{N2}$ by sudden ordering of the kagome spins (Figure \ref{fgr:4}b), which account for most of the integer peak intensity and thus display long-range order. The specific heat also begins to drop dramatically below $T_\mathrm{N2}$. The interlayer \ce{Cu$^{2+}$} correlations are probably affected by the random occupancy of the interlayers at low temperatures. Even below the structural transition temperature, the kagome lattice remains nearly perfect, and the strong interactions between kagome \ce{Cu$^{2+}$} moments can drive longer range correlations.

\subsection{\label{sec:ZnBarlowiteStructure}Structure and magnetism of the Zn-doped quantum spin liquid compositions: ideal kagome planes and robust QSL behavior}

Substituting Zn into barlowite can yield polycrystalline samples with no magnetic order---the first step to a QSL candidate material.\cite{Feng2017} We synthesized sizable single crystals of Zn-substituted barlowite and determined their crystal structure along with that of polycrystalline samples. Chemical analysis (ICP-AES) indicated the crystals' composition is \ce{Cu$_{3.44}$Zn$_{0.56}$(OH)6FBr} (denoted \textbf{Zn$_{0.56}$}). A deuterated polycrystalline sample (denoted \textbf{Zn$_{0.95}$}) has nearly a full equivalent of Zn, which is comparable to (but higher than) that reported in herbertsmithite.\cite{Freedman2010} The deuteration, which is important for neutron scattering measurements, does not have an appreciable effect on the magnetic behavior.

Neither compound demonstrates symmetry lowering down to the lowest temperatures measured, leading to perfect, undistorted kagome lattices. For polycrystalline \textbf{Zn$_{0.95}$}, neutron and synchrotron X-ray data down to $T = 3$ K were co-refined in space group \textit{P}6$_3$/\textit{mmc}, and single crystalline \textbf{Zn$_{0.56}$} was measured with synchrotron X-ray diffraction down to $T = 90$ K (Supplementary Figures 1 and 7--10; Supplementary Tables 2, 4--7, and 12--14). While the interlayer \ce{Zn$^{2+}$}/\ce{Cu$^{2+}$} in herbertsmithite occupy the same shared crystallographic site, in Zn-substituted barlowite the \ce{Zn$^{2+}$} lies on a centered $D_{3h}$ position in perfect trigonal prismatic coordination while the \ce{Cu$^{2+}$} is equally disordered over the three off-center $C_{2v}$ sites (Figure \ref{fgr:5}a).  Rietveld refinements of the neutron data for \textbf{Zn$_{0.95}$} support this assignment, and it is consistent with the different Jahn-Teller activity of these ions.  We determine that the average structure is hexagonal; however, we cannot exclude subtle local symmetry lowering or lattice distortions that evade X-ray crystallographic methods, which are sensitive to the bulk long-range order.

Both compositions of Zn-substituted barlowite display a lack of magnetic order consistent with a QSL ground state, as shown in Figure \ref{fgr:5}b. The susceptibility of single crystalline \textbf{Zn$_{0.56}$} is higher than that of polycrystalline \textbf{Zn$_{0.95}$} and herbertsmithite due to the higher concentration of magnetic impurities on the interlayer ($\sim$44\% \ce{Cu$^{2+}$}). Interestingly, these results imply that even $\sim$50\% \ce{Zn$^{2+}$} on the interlayer is sufficient to push the ground state towards a QSL. In herbertsmithite, the interlayer sites form a cubic superlattice, although only 15\% are occupied by \ce{Cu$^{2+}$}---which is below the percolation limit of 0.31.\cite{Han2016b,Wang2013} Here, the amount of magnetic \ce{Cu$^{2+}$} impurities in \textbf{Zn$_{0.56}$} is well above this percolation limit, but long-range order is not observed.

Synthesizing two samples of Zn-substituted barlowite with different amounts of Zn allows us to quantify the impurity (interlayer \ce{Cu$^{2+}$}) contribution to the total susceptibility. By removing the impurity contribution, an estimate of the intrinsic susceptibility of the barlowite kagome lattice is revealed (Figures \ref{fgr:5}b). The susceptibility of herbertsmithite (which has approximately 15\% \ce{Cu$^{2+}$} interlayer impurities\cite{Freedman2010}) nearly overlays that of the impurity-subtracted \textbf{Zn$_{0.56}$} at low temperatures, although they diverge slightly above $T \approx 50$ K. \textbf{Zn$_{0.95}$} is very close to ``pure,'' and its largest deviations from impurity-subtracted \textbf{Zn$_{0.56}$} occur at the lowest temperatures, below $T \approx 20$ K. The inverse susceptibility of the impurity contribution (Figure \ref{fgr:5}b, inset) shows linear Curie-Weiss behavior below $T = 50$ K with a small AF $\Theta = -3$ K, similar to herbertsmithite.\cite{Bert2007,Han2012b} Assuming a $\mu_\mathrm{eff}=1.8 \upmu_\mathrm{B}$ as observed in the other samples produced 0.45 \ce{Cu$^{2+}$}s, consistent with the 0.39 \ce{Cu$^{2+}$}s from chemical analysis. Curie-Weiss fits indicate that impurity-subtracted \textbf{Zn$_{0.56}$} has the most negative $\Theta = -253$ K, signaling the highest degree of magnetic frustration in this family (Supplementary Table 15; Supplementary Figures 13--14).  

In single crystal \textbf{Zn$_{0.56}$}, the low-temperature magnetic response \textit{M} vs. $\mu_\mathrm{0}H$ is anisotropic (Figure \ref{fgr:5}c). The \textit{ab} plane magnetization is higher than the \textit{c}-axis magnetization, consistent with the trend in barlowite \textbf{2} that reflects the anisotropy of the interlayer \ce{Cu$^{2+}$} moments. However, no hysteresis is observed, which further indicates the lack of magnetic ordering or freezing. The magnitude of the magnetization is approximately half that of barlowite \textbf{2}, consistent with the significant Zn substitution of $x = 0.56$ on the interlayer. Curie-Weiss fits on oriented single crystalline barlowite \textbf{2} and \textbf{Zn$_{0.56}$} (Supplementary Figure 15 and Supplementary Table 16) confirm the clear magnetic anisotropy observed in the \textit{M} vs. $\mu_\mathrm{0}H$ curves, with $\chi_\mathrm{c} > \chi_\mathrm{ab}$ at high temperatures, similar to herbertsmithite, which indicates the presence of easy-axis exchange anisotropy.\cite{Han2011,Han2012b}

Figures \ref{fgr:5}d and e show the specific heat plotted as $C/T$ vs. $T$ of \textbf{Zn$_{0.95}$} and \textbf{Zn$_{0.56}$} from $T = 0.1$--$10$ K; $C$ vs. $T$ and $T/(\mu_\mathrm{0}H)$ are plotted in Supplementary Figure 17. The lack of a sharp anomaly in both samples (similar to herbertsmithite) is consistent with the absence of magnetic order. The low-temperature upturn in both samples can be attributed to a nuclear Schottky anomaly from Cu. \textbf{Zn$_{0.95}$} exhibits a broad feature below $T\approx 3$ K that shifts to higher temperatures and broadens in an applied field, similar to observations in herbertsmithite.\cite{Helton2007}  \textbf{Zn$_{0.56}$} also exhibits a broad peak in $C/T$, where the area under the curve shifts to higher temperature with increasing field.  Interestingly, the area under the $C/T$ curve, which measures the change in entropy, is roughly similar in both \textbf{Zn$_{0.56}$} and \textbf{Zn$_{0.95}$} samples. Hence the degrees of freedom that give rise to this broad peak do not simply scale with the number of interlayer impurities. Supplementary Figure 18 shows the suppression of magnetic order across the barlowite family in $C$ vs. $T$. For Zn-substituted compositions \ce{Cu$_{4-x}$Zn$_{x}$(OH)6FBr} with $x\geq0.5$, the quantum spin liquid signatures in the bulk properties appear surprisingly similar.

\section{\label{sec:Discussion}Discussion}

Comparing the three flavors of the modulated kagome lattice (orthorhombic distorted, pinwheel, and uniform) leads to new insights about the competition between the quantum ground states. Single crystal growth and combined high-resolution synchrotron x-ray and neutron diffraction (structural and magnetic) were key advances in obtaining these results. A summary of the interplay between the quantum spin liquid, valence bond crystal, and magnetic order ground states in the barlowite family is shown in Figure \ref{fgr:8}. First, we find that QSL behavior is surprisingly robust in \textbf{Zn$_{0.56}$} despite a substantial number of interlayer impurities. The magnetic susceptibility of the kagome moments is remarkably similar to herbertsmithite, even though we deduce that the interlayer magnetic impurities occupy lower symmetry sites in both \textbf{Zn$_{0.56}$} and \textbf{Zn$_{0.95}$}. While the strong FM interlayer coupling to the kagome spins is responsible for the ordering in barlowite \textbf{2}, no FM correlations are evident in \textbf{Zn$_{0.56}$} (nor \textbf{Zn$_{0.95}$}), even though 44\% of the interlayer sites contain Cu$^{2+}$. A strong relative distinction of our Zn-substituted compositions is that the kagome lattice is not modulated, preferring to stabilize the QSL. Hence, it appears that once spins participate in the QSL wavefunction, they resist forming substantial correlations with impurity moments.

In barlowite \textbf{2}, the subtle pattern of bond modulations is shown by DMRG calculations to prefer a VBC ground state---which has long been suspected to be close in energy to the QSL state. An interesting pinwheel VBC state has been previously observed in \ce{Rb2Cu3SnF12};\cite{Matan2010} however, that material has bond deformations that are over ten times larger than those we observe in barlowite \textbf{2}. Here, we find that even a less than one percent modulation of the bonds results in a VBC ground state.  We note that in barlowite \textbf{2}, the VBC does not order spontaneously but is rather pinned by the preexisting lattice distortion. The additional coupling to the interlayer spins eventually induces a \textit{q=0}-type spin order, which is also known to be close in energy to the QSL phase. The \textit{q=0} structure is adjacent to the QSL ground state in the phase diagram when second neighbor AF Heisenberg coupling and/or a DM interaction is included.\cite{Messio2010,Huh2014,Suttner2014}

Barlowite \textbf{2} has a new Hamiltonian that has not been studied as yet, consisting of minute deviations from the uniform kagome lattice. We may rationalize the magnetic transitions upon cooling in the following way. At high temperature, the spin fluctuations are likely very similar to those in the QSL phase (Zn-substituted barlowite with $x\gtrsim0.5$). Below the structural phase transition, the distinction between $J$ and $J'$ onsets gradually and likely selects VBC correlations at lower temperatures, although they are obscured by competing thermal effects. Below $T_\mathrm{N1} = 10$ K, the interlayer spins begin to exhibit short-range order with a growing moment, which acts as a staggered internal magnetic field; once it grows strong enough to overcome the VBC correlations, it drives \textit{q=0} order at $T_{N2} = 6$ K. Therefore in barlowite {\bf 2}, we can identify two specific competing ground states---pinwheel VBC and pinwheel \textit{q=0}---which are adjacent to the QSL ground state of the neighboring Zn-substituted compounds.

Clearly, different synthetic routes significantly affect the morphology and low-temperature properties of barlowite, and this is a new lever to explore this quantum material.\cite{Smaha2018}  The differences point to an additional level of complexity in the synthesis of Cu oxysalt minerals and calls for a re-examination of the existing characterization of this class of materials, which also includes claringbullite\cite{fejer1977,burns1995a,feng2019} and clinoatacamite,\cite{Jambor1996,Malcherek2017,Wills2008} among others. The two distinct structures below $T \approx 265$ K in the two variants of barlowite demonstrate a clear structure-properties relationship with the low-temperature magnetism.

The strong synergy between theory and experiment in $S=\frac{1}{2}$ kagome systems makes comparisons between different kagome QSL candidates timely and intriguing.\cite{Balents2010,Norman2016,Mendels2016,Savary2017} Zn-substituted barlowite is only the second $S=\frac{1}{2}$ kagome QSL compound after herbertsmithite; the main differences between these compounds is the slightly off-center location of the interlayer moments and the stacking sequence of the kagome layers. Comparing herbertsmithite and Zn-substituted barlowite not only furthers our understanding of the low energy behavior caused by slight imperfections (such as impurities, lattice distortions, or magnetic interactions beyond the idealized models) which are inherent in all QSL materials, but also illuminates the rich rivalry of ground states that compete with the exotic QSL ground state. The ability to realize specific patterns of modulated bonds on the kagome lattice opens the door to investigate possible new classes of modulated QSL states that break the lattice symmetry.\cite{ClarkBryan2013}

In summary, we have made the first large single crystals of two compounds, which allows us unique access to their physics. Two resulting highlights are: 1) our discovery of VBC physics on a patterned kagome lattice close to the QSL, and 2) the robustness of QSL behavior on the uniform kagome lattice even in the presence of many interlayer impurities. We have additionally performed DMRG calculations to predict the ground state physics of the lattices that we determined. Finally, the quality of our structural analysis is unprecedented for the barlowite family, which is extremely important to understand these delicate phases. Convincing proof of new quantum phases requires strong evidence of sample quality, which we provide here. The universal magnetic behavior that we find in the two QSL compositions is exciting as the kagome quantum spin liquids are one of only a few physical systems that are believed to have long-range quantum entanglement.

\section{Methods}
  
All hydrothermal reactions were performed in PTFE-lined stainless steel autoclaves, and the products were recovered by filtration and washed with DI \ce{H2O}.  The reactions yielding barlowite \textbf{2} and \textbf{Zn$_{0.56}$} crystals also yielded polycrystalline \ce{LiF}, which was removed by sonication in acetone. Barlowite \textbf{1}\cite{Smaha2018} was synthesized by mixing \ce{Cu2(OH)2CO3} (Alfa, Cu 55\%), \ce{NH4F} (Alfa, 96\%), \ce{HBr} (Alfa, 48\% wt), and 36 mL DI \ce{H2O} (EMD Millipore) or \ce{D2O} (Aldrich, 99.9\%) in a 45 mL autoclave, which was heated over 3 h to 175 $^{\circ}$C and held for 72 h before being cooled to room temperature over 48 h.  Protonated barlowite \textbf{2}\cite{Smaha2018} was synthesized by mixing \ce{CuF2} (BTC, 99.5\%), \ce{LiBr} (Alfa, 99\%), and 15 mL DI \ce{H2O} in a 23 mL autoclave, which was heated over 3 h to 175 $^{\circ}$C and held for 72 h, then cooled to 80 $^{\circ}$C over 24 h. It was held at 80 $^{\circ}$C for 24 h before being cooled to room temperature over 12 h. Deuterated barlowite \textbf{2} was prepared by combining a stoichiometric mixture of \ce{CuF2} (6 mmol) and \ce{LiBr} (10.5 mmol) with 15 mL \ce{D2O} in a 23 mL autoclave, which was then heated to over 3 h to 175 $^{\circ}$C and held for 72 h, then cooled to 80 $^{\circ}$C over 48 h. It was held at 80 $^{\circ}$C for 96 h before being cooled to room temperature over 24 h. \textbf{Zn$_{0.56}$} crystals were grown using 14 excess equivalents of \ce{ZnF2}. Adding \ce{ZnF2} in excess of this resulted in phase separation of the products. \ce{CuF2} (1.8 mmol), \ce{ZnF2} (9 mmol; Alfa, 99\%), and \ce{LiBr} (21 mmol) were sealed in a 23 mL autoclave with 15 mL DI \ce{H2O}.  This was heated over 3 hours to 215 $^{\circ}$C and held for 72 hours, then cooled to 80 $^{\circ}$C over 48 hours. It was held at 80 $^{\circ}$C for 72 hours before being cooled to room temperature over 24 hours. \textbf{Zn$_{0.95}$} and herbertsmithite were synthesized using \ce{Cu2(OH)2CO3}, \ce{NH4F}, and \ce{ZnBr2} (BTC, 99.999\%) or \ce{ZnCl2} (Alfa, 99.999\%) in a 23 mL autoclave with 10 mL DI \ce{H2O} or \ce{D2O}. This was heated over 3 h to 210 $^{\circ}$C and held for 24 h before being cooled to room temperature over 30 h.\cite{Shores2005,Han2011,Smaha2018}

Synchrotron single crystal diffraction (SCXRD) experiments were conducted at NSF's ChemMatCARS beamline 15-ID at the Advanced Photon Source (APS), Argonne National Laboratory, using a Bruker D8 diffractometer equipped with a PILATUS3 X CdTe 1M detector or a Bruker APEX II detector. Data sets were collected using a wavelength of 0.41328 \AA. Additional data sets were collected at beamline 12.2.1 at the Advanced Light Source (ALS), Lawrence Berkeley National Laboratory, using a Bruker D85 diffractometer equipped with a Bruker PHOTON II detector and a wavelength of 0.7288 \AA. The data were integrated and corrected for Lorentz and polarization effects using SAINT and corrected for absorption effects using SADABS.\cite{BrukerAXSSoftwareInc2016} The structures were solved using intrinsic phasing in APEX3 and refined using the SHELXTL and OLEX2 software.\cite{Sheldrick2015,Dolomanov2009} Hydrogen atoms were inserted at positions of electron density near the oxygen atom and were refined with a fixed bond length and an isotropic thermal parameter 1.5 times that of the attached oxygen atom. Thermal parameters for all other atoms were refined anisotropically. 

Synchrotron powder X-ray diffraction (PXRD) data were collected at APS beamline 11-BM using wavelengths of 0.412728, 0.412702, and 0.457676 \AA. Samples were measured in Kapton capillaries; crystalline samples were crushed into a powder. Rietveld refinements were performed using GSAS-II.\cite{Toby2013} Atomic coordinates and isotropic atomic displacement parameters were refined for each atom; site occupancy was also refined for the interlayer site when appropriate. Hydrogen was excluded.

Neutron powder diffraction (NPD) data were collected using the BT-1 32 detector neutron powder diffractometer at the Center for Neutron Research (NCNR) at the National Institute of Standards and Technology (NIST). A Ge(311) monochromator with a 75$^{\circ}$ take-off angle, wavelength of 2.0775 \AA, and in-pile collimation of 60$'$ was used. Data were collected from 1.3--166.3$^{\circ}$ 2$\Theta$ with a step size of 0.05$^{\circ}$. The samples were loaded in vanadium cans. Rietveld co-refinements of PXRD and NPD data were performed using GSAS-II. Atomic coordinates and isotropic atomic displacement parameters were refined for each atom, including deuterium; site occupancy was also refined for the interlayer site when appropriate.

DC magnetization and AC susceptibility measurements were performed on a Quantum Design Physical Properties Measurement System (PPMS) from $T = 2 - 350$ K in applied fields up to $\mu_\mathrm{0}H = 9$ T. AC susceptibility was measured with a 1.5 Oe drive current at a frequency of 10,000 Hz. DC susceptibility was measured on a Quantum Design Magnetic Property Measurement System (MPMS) from $T = 2 - 300$ K in applied fields up to $\mu_\mathrm{0}H = 7$ T.  Heat capacity measurements were performed in the PPMS with a dilution refrigerator on either a pressed single pellet of powder mixed with Ag powder in a 1:2 mass ratio or on a single crystal affixed to a sapphire platform using Apiezon-N grease. Ag powder was measured separately, and its contribution was subtracted.

Elastic neutron scattering measurements were performed on aligned single crystals of deuterated barlowite \textbf{2} with a total mass of 72.8 mg (266 coaligned crystals) at beamline HB-1A, the Fixed-Incident-Energy Triple-Axis Spectrometer (FIE-TAX), at the High Flux Isotope Reactor (HFIR), Oak Ridge National Laboratory. The incident and final energies were fixed at 14.656 meV, with horizontal collimation of 40$'$-40$'$-sample-80$'$-open, and a liquid helium cryostat was used to reach the base temperature of $T = 1.6$ K. Additional measurements were taken at the Spin Polarized Inelastic Neutron Spectrometer (SPINS) and the BT-7 Double Focusing Thermal Triple Axis Spectrometer\cite{BT7} at the NCNR with a total sample mass of 226.7 mg (1115 coaligned crystals). For SPINS, the incident and final energies were fixed at 5 meV, with horizontal collimation of guide-open-sample-80$'$-open. For BT-7, the incident and final energies were fixed at 14.7 meV, with horizontal collimation of open-80$'$-sample-80$'$-120$'$. Base temperatures of $T = 4$ K (for one experiment at SPINS) and $T = 2.7$ K (BT-7 and another experiment at SPINS) were achieved using closed cycle refrigerators. The sample mosaic was 3.00$^\circ$ to 3.33$^\circ$ (FWHM) for all experiments, depending on the orientation. More details of the magnetic model analysis can be found in the Supplementary Material.

We employ the standard DMRG method\cite{White1992} to investigate the ground state properties of the S=$\frac{1}{2}$ antiferromagnetic $J-J'$ Heisenberg model on the kagome lattice. We take the lattice geometry to be cylindrical and lattice spacing to be unity. The boundary condition of the cylinder is periodic around the cylinder but open along the cylinder. We focus on cylinders of width $L_y=4$, i.e., 4 unit cells or 8 sites around the cylinder. We keep around $m=4000$ states in each DMRG block with a typical truncation error $\epsilon\sim 10^{-6}$ and perform more than 16 sweeps, which leads to excellent convergence of our results.

\section{Data Availability}
The X-ray and neutron crystallographic coordinates for structures reported in this study have been deposited at the Cambridge Crystallographic Data Centre (CCDC), under deposition numbers 1898751--4, 1898756, 1899242--3, 1899246, 1899248, 1899250, 1899252, 1899365, 1899367, 1899369, 1899370, and 1922923. These data can be obtained free of charge from The Cambridge Crystallographic Data Centre via www.ccdc.cam.ac.uk/data\_request/cif.  Raw data were generated at the Advanced Photon Source (beamlines 15-ID and 11-BM) and Advanced Light Source synchrotron radiation facilities (beamline 12.2.1) as well as the High Flux Isotope Reactor (beamline HB-1A) and NIST Center for Neutron Research neutron sources (beamlines BT-1, BT-7, and SPINS). Data supporting the findings of this study are available from the corresponding authors upon reasonable request.  

\section{Code Availability}
The magnetic structure refinement code is available from the corresponding authors upon reasonable request.  

\begin{acknowledgments}
The work at Stanford and SLAC was supported by the U.S. Department of Energy (DOE), Office of Science, Basic Energy Sciences (BES), Materials Sciences and Engineering Division, under Contract No. DE-AC02-76SF00515. A portion of this research used resources at the High Flux Isotope Reactor, a DOE Office of Science User Facility operated by the Oak Ridge National Laboratory. We acknowledge the support of the National Institute of Standards and Technology, U. S. Department of Commerce, in providing the neutron research facilities used in this work. This research used resources of the Advanced Light Source, which is a DOE Office of Science User Facility under contract No. DE-AC02-05CH11231. Use of the Advanced Photon Source, an Office of Science User Facility operated for the U.S. DOE Office of Science by Argonne National Laboratory, was supported by the U.S. DOE under Contract No. DE-AC02-06CH11357. NSF's ChemMatCARS Sector 15 is supported by the Divisions of Chemistry (CHE) and Materials Research (DMR), National Science Foundation, under grant number NSF/CHE-1834750.  Use of the PILATUS3 X CdTe 1M detector is supported by the National Science Foundation under grant number NSF/DMR-1531283. Part of this work was performed at the Stanford Nano Shared Facilities (SNSF), supported by the NSF under award ECCS-1542152. R.W.S. was supported by the Department of Defense (DoD) through the NDSEG Fellowship Program and by an NSF Graduate Research Fellowship (DGE-1656518). We thank S. Lapidus for assistance at APS beamline 11-BM; C.M. Brown for assistance at NCNR beamline BT-1; I.R. Fisher for use of the MPMS and dilution refrigerator; A.T. Hristov for help with the dilution refrigerator; and S. Raghu, T. Senthil, and G. Chen for helpful discussion. The identification of any commercial product or trade name does not imply endorsement or recommendation by the National Institute of Standards and Technology.

\end{acknowledgments}

 \section{Competing Interests}
 The Authors declare no competing financial or non-financial interests.

\section{\label{sec:Contributions}Author Contributions}
 R.W.S. and Y.S.L. conceived the study, interpreted the data, and wrote the manuscript with contributions and comments from all authors. R.W.S. synthesized barlowite \textbf{2} and all Zn-substituted barlowite samples; performed and analyzed X-ray and neutron diffraction measurements; and performed and analyzed magnetic susceptibility and heat capacity measurements. W.H. synthesized barlowite \textbf{1} and performed and analyzed neutron scattering measurements. J.M.J., C.J.T., and J.W. aided at beamtimes and with data analysis. Y.-F.J. and H.C.J. performed numerical calculations. J.P.S. synthesized herbertsmithite. S.G.W., Y.-S.C., S.J.T, A.A.A., Y.Z., G.X, and J.W.L. aided at beamtimes.

\end{flushleft}

\section{Figure Legends}

\begin{figure*}[htbp]
\includegraphics[width=14cm]{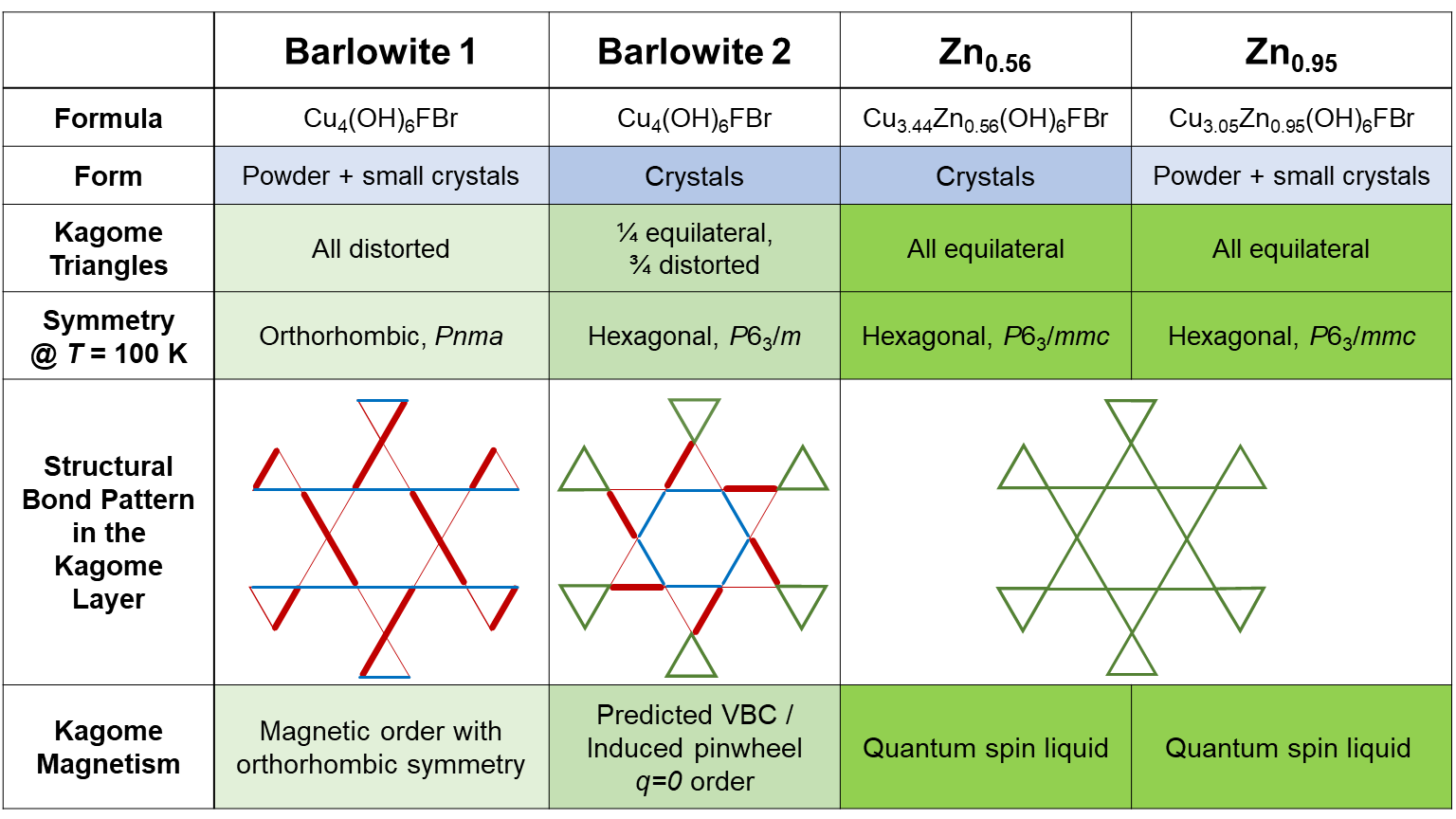}
\caption{\textbf{Quantum ground states stabilized by modulations of the kagome lattice.} Structural and magnetic characteristics are tabulated. The second to last row shows the magnetic interactions on the kagome planes of barlowite \textbf{1}, barlowite \textbf{2}, and Zn-substituted barlowite. The thickness of the line indicates bond strength as expressed through its Cu--O--Cu bond angle. In barlowite {\bf 2} the variation in angle of $\pm 0.8^{\circ}$ is quite small, so the differences in magnetic exchange are also likely small. In \textbf{Zn$_{0.56}$} and \textbf{Zn$_{0.95}$}, the kagome motif consists only of equilateral triangles (green); however, in barlowite \textbf{2} the kagome motif is expanded due to a mix of equilateral (green) and distorted (red/blue) triangles. On the other hand, barlowite \textbf{1} is more distorted, containing only distorted triangles.}
\label{fgr:table}
\end{figure*}

\begin{figure*}[htbp]
\includegraphics[width=14cm]{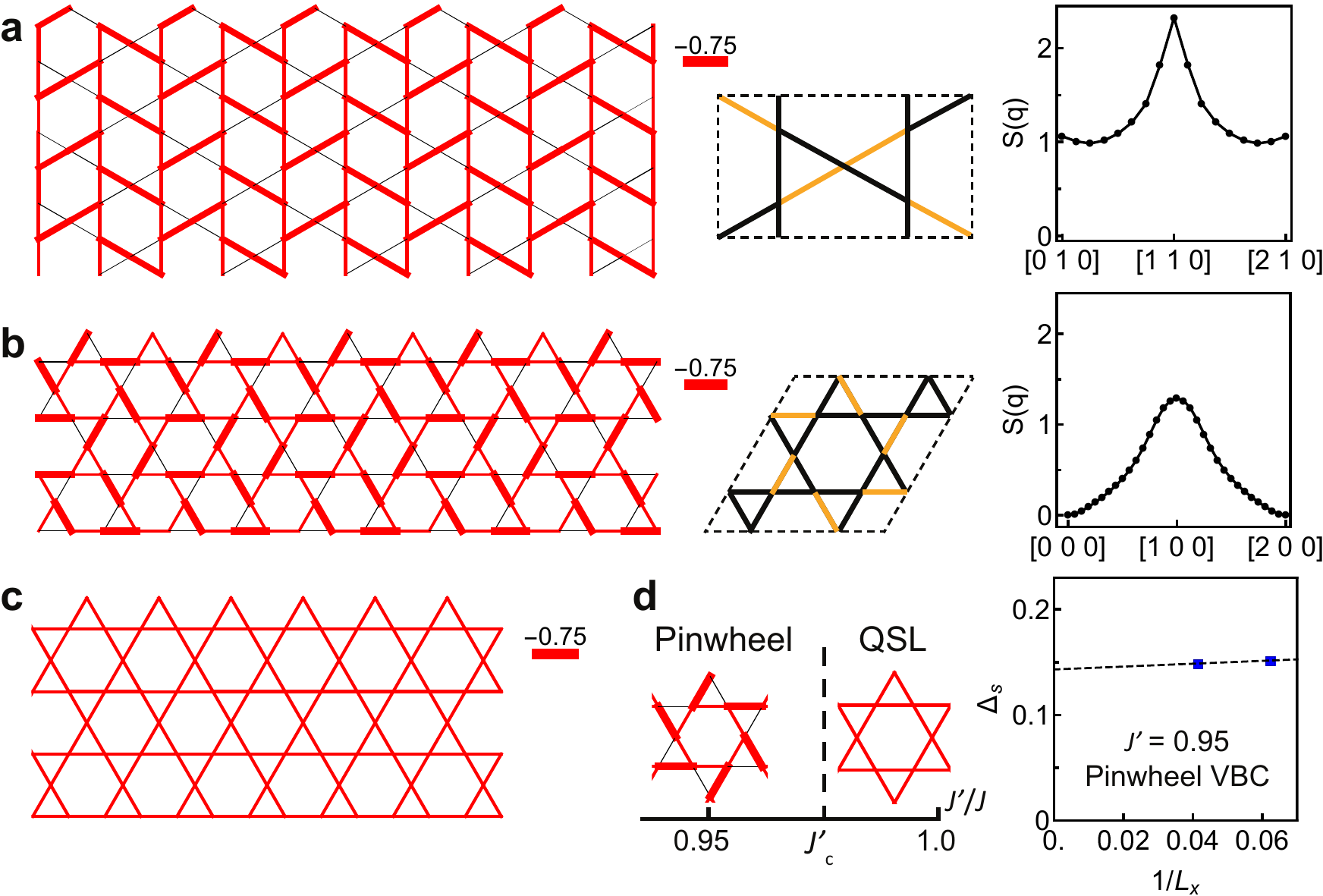}
\caption{\textbf{Numerical simulations predicting three rival ground states.} \textbf{a} Illustration of a simplified $J-J'$ Heisenberg model for barlowite \textbf{1}. The 6-site unit cell is in the middle panel, and the bond pattern for $J'=0.8J$ is shown in the left panel on a cylinder consisting of $16\times 4$ unit cells (192 sites) where the bond thickness denotes the absolute value of $\left\langle {\bf S}_i \cdot{\bf S}_j \right\rangle$. The corresponding spin structure factor $S(q)$ in the right panel has a sharp peak at momentum [1 1 0]. \textbf{b} Illustration of a simplified $J-J'$ Heisenberg model for barlowite \textbf{2}. The 12-site unit cell is in the middle panel, and the bond pattern for $J'=0.95J$ is given in the left panel. The cylinder consists of  $24 \times 4$ unit cells (288 sites). The spin structure factor $S(q)$ in the right panel has a broad peak at momentum [1 0 0].  Here, the yellow lines denote bonds with interaction $J'$, while the black lines are bonds with interaction $J$. \textbf{c} The isotropic Heisenberg model with $J'=J$ and the corresponding uniform bond pattern in the QSL phase. The cylinder consists of $16 \times 4$ unit cells. \textbf{d} Schematic phase diagram of the $J-J'$ Heisenberg model as a function of $J'/J$, where the dashed line $J'_c=0.998(1)$ denotes the phase boundary between the pinwheel VBC and QSL phases. The spin gap $\Delta_s$ of the $J'=0.95J$ point in the pinwheel VBC phase is plotted in the right panel as a function of the inverse diameter of the cylinder. Error bars represent the confidence interval in the fitting procedure corresponding to $\pm\sigma$ ($\sigma$ is the standard deviation).}
\label{fgr:3}
\end{figure*}

\begin{figure*}[htbp]
\includegraphics[width=16cm]{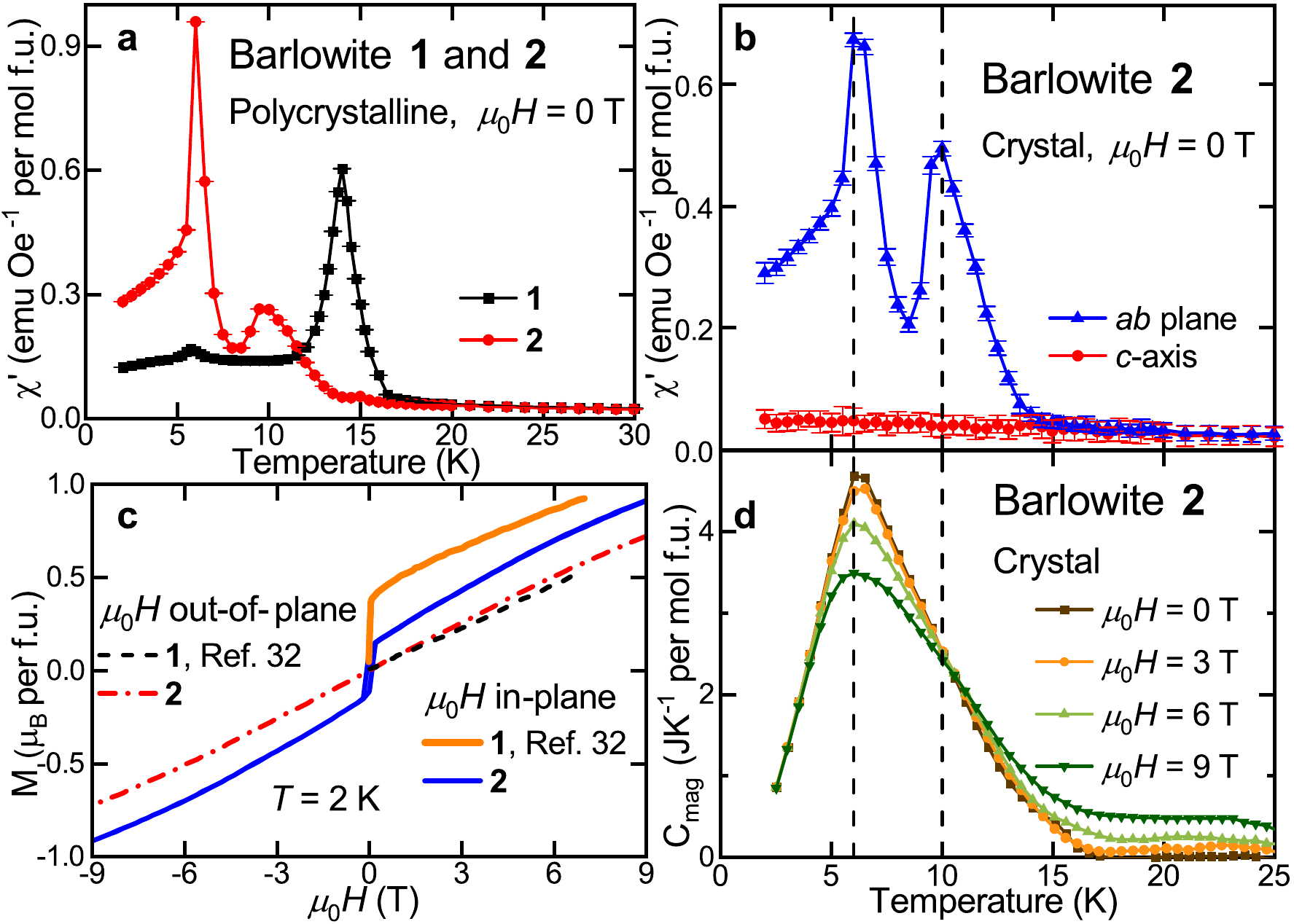}
\caption{\textbf{Magnetic phase transitions in barlowite 2.} \textbf{a} The real part of the AC susceptibility ($\chi$') of polycrystalline, orthorhombic barlowite \textbf{1} and a collection of single crystals of hexagonal barlowite \textbf{2}. \textbf{b} The real part of the AC susceptibility ($\chi$') of a 2.0 mg single crystal of barlowite \textbf{2} aligned in the \textit{ab} plane and along the \textit{c}-axis. \textbf{c} Aligned magnetization ($M$) as a function of applied field at $T = 2$ K for single crystals of barlowite \textbf{2} and \textbf{1} (from Ref. \cite{Han2016}). \textbf{d} Heat capacity associated with the magnetic transitions ($C_\mathrm{mag}$) of barlowite \textbf{2} calculated as $C-C_\mathrm{bg}$ (see Supplementary Figure 16). The dashed black lines denote the magnetic transitions of barlowite \textbf{2}: $T_\mathrm{N1} = 10$ K and $T_\mathrm{N2} = 6$ K. Error bars represent the confidence interval in the fitting procedure corresponding to $\pm\sigma$ ($\sigma$ is the standard deviation). Note:  1 emu (mol Oe)$^{-1}= 4 \pi 10^{-6}$ m$^3$ mol$^{-1}$. }
\label{fgr:barl}
\end{figure*}

\begin{figure*}[htbp]
\includegraphics[width=15cm]{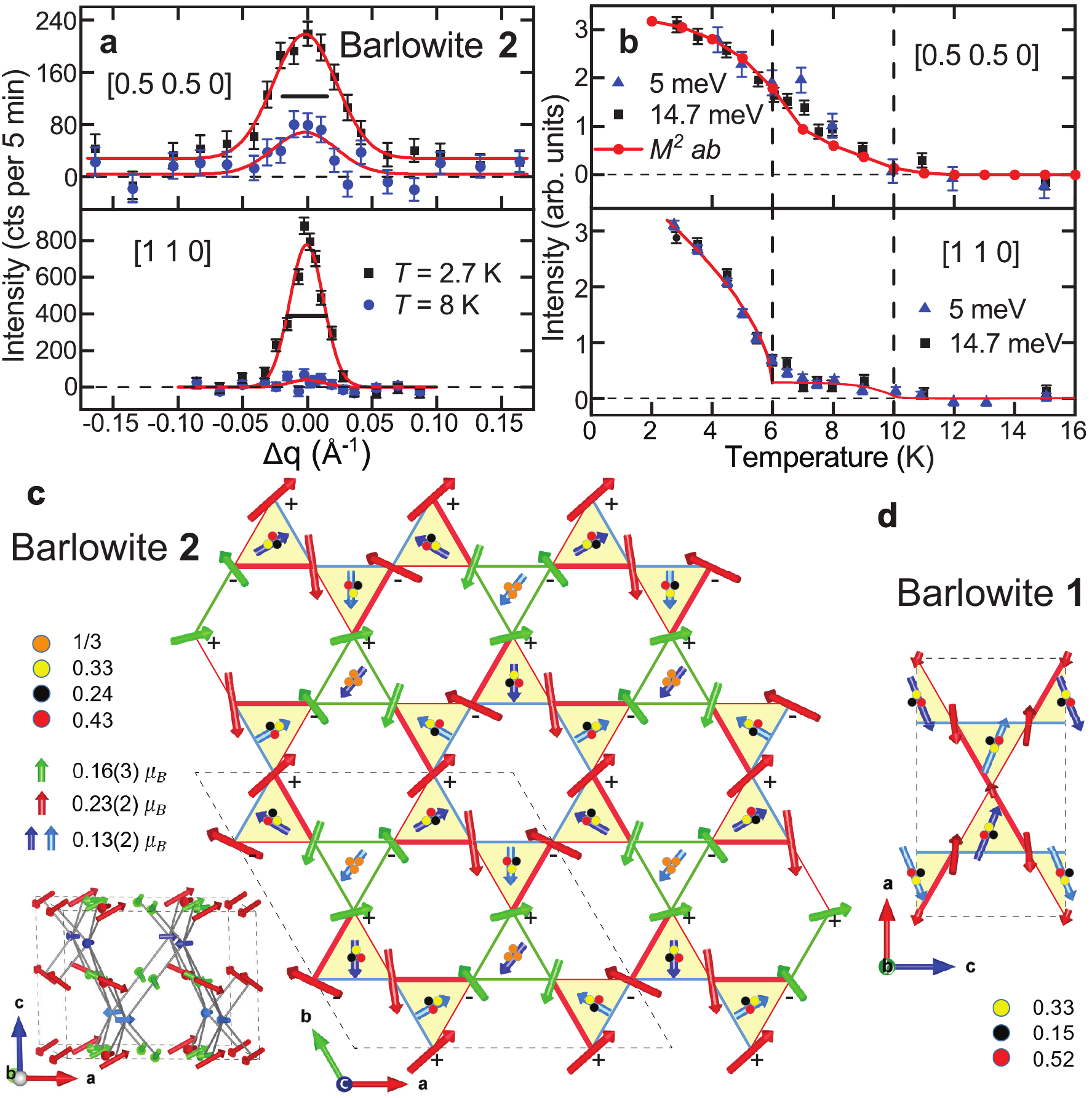}
\caption{\textbf{Magnetic state of barlowite 2 elucidated by magnetic neutron diffraction on a single crystal.} \textbf{a} Longitudinal scans through the [0.5 0.5 0] and [1 1 0] magnetic Bragg peaks measured at $T = 2.7$ and $8$ K with $E_\mathrm{i} = 5$ meV at SPINS, where the $T = 20$ K data was subtracted as background. The red lines are Gaussian fits. The horizontal bars indicate the experimental resolution, and the error bars represent one standard deviation. \textbf{b} Temperature dependences of the peak intensities of the [0.5 0.5 0] and [1 1 0] magnetic Bragg peaks measured at $E_\mathrm{i} = 5$ meV (SPINS) and $E_\mathrm{i} = 14.7$ meV (BT-7). The top panel also displays the squared magnetization ($M^2$) of a single crystal in the \textit{ab} plane at $\mu_\mathrm{0}H = 0.005$ T. Below $T = 6$ K, the red line in the bottom panel is a power-law fit from $T = 2$ K to $11$ K with critical exponent $\beta = 0.30(3)$ and $T_\mathrm{N} = 6$; above $T = 6$ K the line is a guide to the eye.  \textbf{c} The pinwheel \textit{q=0} magnetic model of barlowite \textbf{2} in the \textit{ab} plane and along \textit{c} (inset). \textbf{d} The magnetic model proposed for orthorhombic barlowite \textbf{1} by Ref. \cite{Tustain2018} overlaid on our barlowite \textbf{1} crystal structure. All occupancy values are taken from SCXRD (see Supplementary Table 3). For \textbf{c} and \textbf{d}, the arrows indicate the sizes and directions of the moments. Dark and light blue spins represent interlayer \ce{Cu$^{2+}$}s in different layers, while red and green spins represent kagome \ce{Cu$^{2+}$}s. Each interlayer motif is visualized by one spin at the weighted center of the sites. The thickness of the lines between kagome spins indicates relative bond strength extracted from the Cu--O--Cu bond angles. The $\pm$ symbols next to the kagome spins denote the directions of the out-of-plane component in the spins. The dashed lines denote the magnetic unit cell.}
\label{fgr:4}
\end{figure*}

\begin{figure*}[htbp]
\includegraphics[width=15cm]{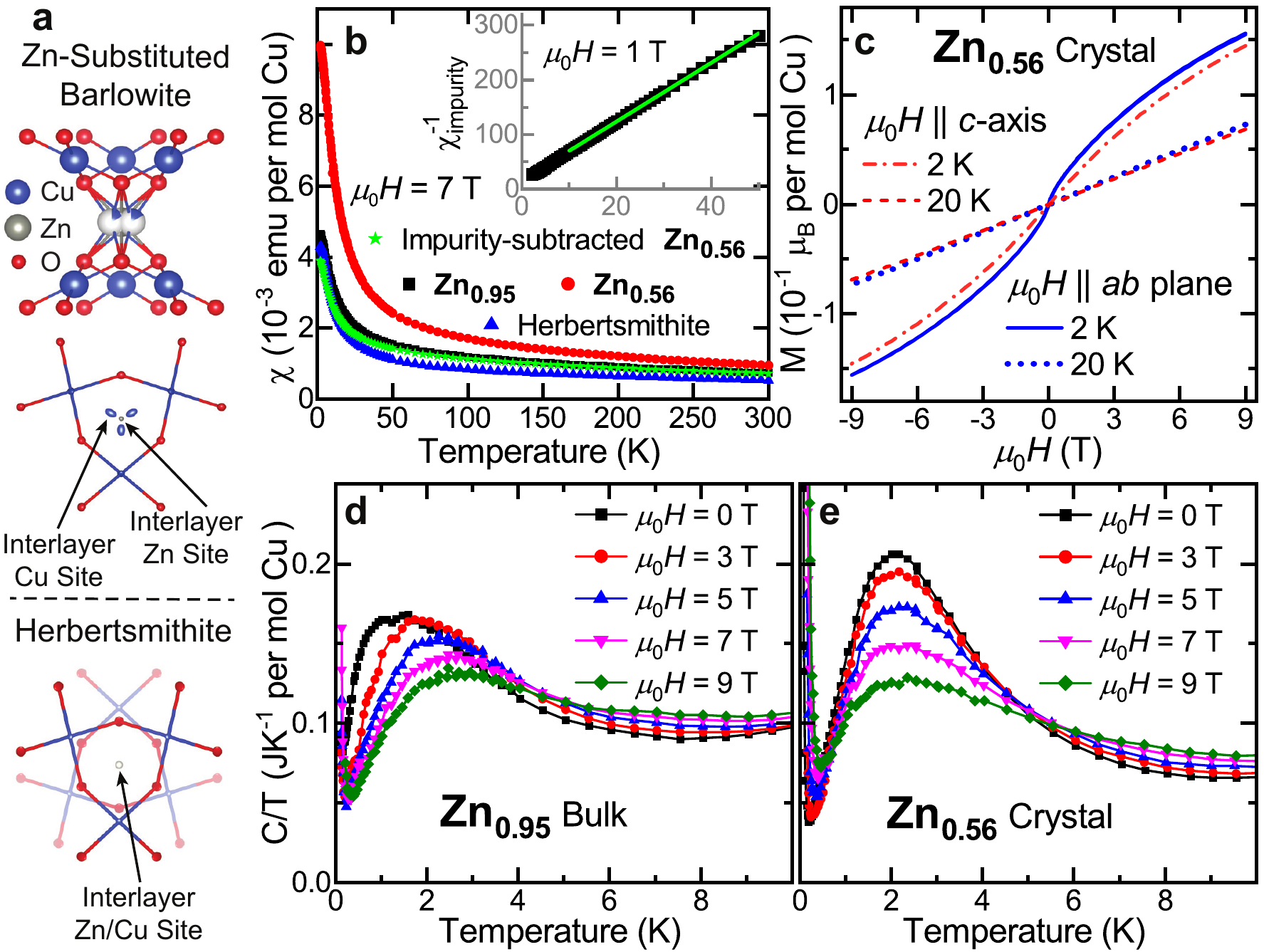}
\caption{\textbf{Structure and quantum magnetism of Zn-substituted barlowite.} \textbf{a} Structures of Zn-substituted barlowite and herbertsmithite showing their interlayer metal coordination environments.  The top-down view  of Zn-substituted barlowite displays the coexistence of the centered ($D_{3h}$) \ce{Zn$^{2+}$} site and the set of triplicated off-center ($C_{2v}$) \ce{Cu$^{2+}$} interlayer sites, visualized in VESTA.\cite{Momma2011} The top-down view of herbertsmithite shows its distinct interlayer coordination: there is only one site for both \ce{Zn$^{2+}$} and \ce{Cu$^{2+}$}, and the kagome layers above and below the interlayer site are rotated by 60$^{\circ}$. In the top-down views, the atoms are displayed as thermal ellipsoids at 90\% probability. F, Br, Cl, and H are not shown. \textbf{b} DC susceptibility ($\chi$) measured in an applied field $\mu_\mathrm{0}H = 7$ T of polycrystalline \textbf{Zn$_{0.95}$}, a collection of single crystals of \textbf{Zn$_{0.56}$}, and polycrystalline herbertsmithite. Also plotted is an estimate of the intrinsic susceptibility of the Zn-substituted barlowite kagome lattice. The amount of \ce{Cu$^{2+}$} magnetic impurities in the herbertsmithite sample is estimated at 15\%.\cite{Freedman2010}  The inset shows the inverse susceptibility of the interlayer \ce{Cu$^{2+}$} magnetic impurity; the line is a Curie-Weiss fit from $T = 10$ K to $50$ K with no diamagnetic correction $\chi_\mathrm{0}$. \textbf{c} Aligned  magnetization as a function of applied field for a 1.4 mg single crystal of \textbf{Zn$_{0.56}$}. \textbf{d} Temperature dependence of \textit{C/T} for \textbf{Zn$_{0.95}$} and \textbf{e} \textbf{Zn$_{0.56}$} in applied fields of $\mu_\mathrm{0}H = 0$--$9$ T. Note: 1 emu (mol Oe)$^{-1}= 4 \pi 10^{-6}$ m$^3$ mol$^{-1}$.}
\label{fgr:5}
\end{figure*}

\begin{figure*}[htbp]
\includegraphics[width=16cm]{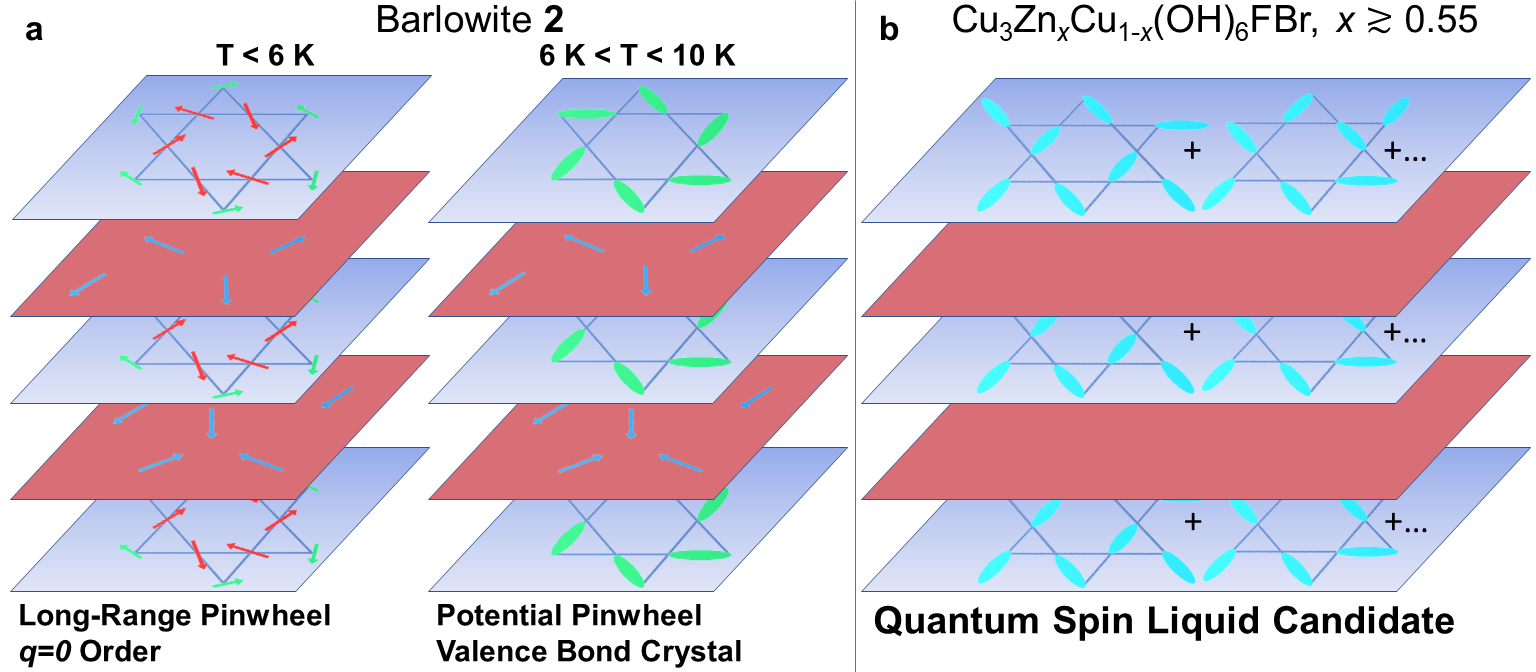}
\caption{\textbf{Competition between magnetic order, valence bond crystal, and quantum spin liquid states in the barlowite system.} \textbf{a} Schematics of the magnetic state of barlowite \textbf{2}; at $T < 6$ K the kagome layer has long-range pinwheel \textit{q=0} order, and at  $6$ K $< T < 10$ K the kagome layer potentially has a pinned pinwheel valence bond crystal state. The interlayer moments are short-range ordered in both temperature ranges. \textbf{b} Schematic of the ground state in Zn-substituted barlowite (\ce{Cu3Zn$_{x}$Cu$_{1-x}$(OH)6FBr}, $x\gtrsim 0.55$). Neither the kagome nor the interlayer spins have appreciable magnetic order down to $T=0.1$ K.}
\label{fgr:8}
\end{figure*}

\end{document}